\documentclass{svjour2}                    
\smartqed  
\usepackage{graphicx}
\usepackage{mathptmx}      
\usepackage{latexsym}

\def\q1{{q^{-1}}}
\def\qq1{{q-q^{-1}}}

\def\dj{{{\cal D}_x}}

\def\nq{{n_{i}}}
\def\be{\begin{equation}}
\def\ee{\end{equation}}
\def\bee{\begin{eqnarray}}
\def\eee{\end{eqnarray}}

\begin{document}
\title{Thermostatistics of deformed bosons and fermions}

\author{A. Lavagno    \and
        P. Narayana Swamy}

\institute{A. Lavagno \at Dipartimento di Fisica, Politecnico di
Torino and INFN - Sezione di Torino, I-10129, Torino, Italy \\
\email{andrea.lavagno@polito.it}           
\and P. Narayana Swamy \at Physics Department, Southern Illinois
University Edwardsville, IL 62026, USA \\
\email{pswamy@siue.edu} }

\date{Received: date / Accepted: date}

\maketitle

\begin{abstract}
Based on the  $q$-deformed oscillator algebra, we study the
behavior of the mean occupation number and its analogies with
intermediate statistics and we obtain an expression in terms of an
infinite continued fraction, thus clarifying successive
approximations. In this framework, we study the thermostatistics
of $q$-deformed bosons and fermions and show that thermodynamics
can be built on the formalism of $q$-calculus. The entire
structure of thermodynamics is preserved if ordinary derivatives
are replaced by the use of an appropriate Jackson derivative (JD)
and $q$-integral. Moreover, we derive the most important
thermodynamic functions and we study the $q$-boson and $q$-fermion
ideal gas in the thermodynamic limit.

\end{abstract}

\section{Introduction}

The founding principle of the quantum theory of many body-systems
and quantum statistical mechanics is based on the spin-statistics
theorem. The contrasting nature of spin one-half particles,
fermions, and spin integer, bosons, affects sensibly the quantum
statistical behavior of many particle systems \cite{spin,berry}.

The power of statistical mechanics lies not only in the derivation
of the general laws of thermodynamics from microscopic theory but
also in determining the meaning of all the thermodynamic functions
in terms of microscopic interparticle interaction. In the recent
literature, a great deal of effort has been devoted to studying
the possible violation of standard statistical behavior derived
from the indistinguishability of identical fermions and bosons.
The investigation of statistical distribution interpolating
between Fermi-Dirac (FD) and Bose-Einstein (BE) distribution
functions has led to many studies of intermediate statistics,
where creation and annihilation operators obey commutation
relation interpolating between bosons and fermions
\cite{gentile,green,greenb1,greenb2,vieira,poli,levine}.
Experimental high precision methods have been developed to verify
small deviations from the Pauli principle
\cite{reines,hilborn,modugno,javorsek,vip}. Implications of such
deviations are reported in atomic, nuclear, high energy physics,
condensed matter, astrophysics and cosmology
\cite{hilborn2,green_hilb,baron,strominger,youm}.

In the above context, describing complex systems, quantum algebra
and quantum groups have been the subject of intense research in
several physical fields. The quantum group $SU_q(2)$ is well-known
to play an important role in classical and quantum integrable
systems and arose in the work on Yang-Baxter equations
\cite{sky,kulish}. The main aspect of quantum theory can be
described by the quantum Heisenberg uncertainty relation and hence
a consequence of $q$-deformation can best be understood in terms
of the  modified uncertainty relation, as revealed in the seminal
works of Biedenharn \cite{bie} and Macfarlane \cite{mac}. The
uncertainty relation, expressed as the commutation relation
between the coordinate and momentum, has the Planck constant,
namely $\hbar$, on the right hand side in the standard quantum
mechanics, is multiplied by a function, $F(q)=\cosh((2n+1)\,(\ln
q)/2)/\cosh((\ln q)/2)$ (where $n$ represents the number of quanta
of oscillators) in the case of the $q$-deformed quantum theory. In
other words, an important consequence of $q$-defor\-ma\-tion can
be stated as a modification of the Planck constant itself.
Furthermore, it was early clear that the $q$-calculus, originally
introduced by Heine\cite{heine} and by Jackson \cite{jack} in the
study of the basic hypergeometric series \cite{gasper}, plays a
central role in the representation of the quantum groups with a
deep physical meaning \cite{cele1,fink,abe,borges1}. A basic
hypergeometric function is a generalization of the ordinary
hypergeometric function, signified by the addition of an extra
parameter $q$, so that in the limit when $q \rightarrow 1$ this
tends to the normal hypergeometric function. These functions have
a range of representation, from the simple basic series to a
generalized basic hypergeometric function. Various ordinary
functions have thus been generalized to introduce the basic
functions with applications in many fields of science.
Furthermore, it is remarkable to observe that the $q$-calculus,
based on the so-called Jackson Derivative (JD) operator and its
inverse operator $q$-integral, is indeed well suited for
describing fractal and multifractal systems. As soon as the system
exhibits a discrete-scale invariance, the natural tool is provided
by Jackson $q$-derivative and $q$-integral, which constitute the
natural generalization of the regular derivative and integral for
discretely self-similar systems \cite{erzan,lava08}.

As a consequence, we expect $q$-calculus to play a central role in
$q$-deformed thermostatistics. It is well-known that the theory of
harmonic oscillators in standard thermodynamics can be modified by
means of the $q$-deformed algebra of the oscillators. This
modifies the thermodynamic distribution function and all the
thermodynamic functions such as the entropy, partition function,
equation of state etc. The $q$-deformation affects the energy
spectrum in a fundamental manner and in this manner the
deformation may be thought of as describing complex ensembles or
real gases, contrary to the standard thermodynamics as the theory
describing ideal gases. The extent of the deformation i.e., the
value of $q$ in reality, is of course to be determined by
confronting the predictions of the theory against experiment and
this has been accomplished so far by means of numerical
calculations. The physical meaning of $q$-deformation can be
stated in terms of the JD i.e., the $q$-deformation corresponds to
difference equations as opposed to continuous derivatives and
continuous differential equations. Several investigators have
studied the equilibrium statistical mechanics of noninteracting
$q$-deformed bosons and fermions and many of the details of the
analysis are well known. It is remarkable that the main premise
consists of demonstrating that the $q$-deformed thermostatistics
can be consistently formulated by replacing the standard ordinary
derivatives of thermostatistics by JD. Of course care must be
exercised in observing that many of the standard rules of calculus
such as the usual Leibniz chain rule are ruled out and in this
manner one formulates what may be termed as the $q$-calculus.
Accordingly, one can formulate the entropy, mean occupation
number, bose condensation etc. and thus the entire
$q$-thermostatistics \cite{pre2000,pre2002,jpa2007}. The main goal
of this paper is to extend our previous investigations employing
consistently $q$-deformed integral (inverse operator of the JD) in
the formulation of the quantum thermodynamics and to study the
behavior of the mean occupation number considering its development
in terms of infinite continued fractions and outlining its
analogies with intermediate statistics.

\section{Deformed algebra and $q$-calculus}

We shall review the principal relations of $q$-oscillators defined
by the $q$-Heisenberg algebra of creation and annihilation
operators introduced by Biedenharn and McFarlane \cite{bie,mac},
derivable through a map from SU$_q$(2). Furthermore, we will
review the main features of the strictly connected $q$-calculus
useful in the present investigation.

The symmetric $q$-oscillator algebra is defined, in terms of the
creation and annihilation operators $c$, $c^\dag$ and the
$q$-number operator $N$, by \cite{ng,chai,lee,song}
\begin{equation}
[c,c]_{\kappa}=[c^\dag,c^\dag]_{\kappa}=0 \; , \ \ \ cc^\dag-
\kappa q c^\dag c =q^{-N} \; ,
\end{equation}
\begin{equation}
[N,c^\dag]= c^\dag \; , \ \ \ [N,c]=-c\; ,
\end{equation}
where the  deformation parameter $q$ is real and $[x,\,
y]_{\kappa}= x y - \kappa y x\,$, where, as before,  $\kappa = 1$
for  $q$-bosons with commutators and $\kappa = -1$ for
$q$-fermions with anticommutators.

Furthermore, the operators obey  the relations
\begin{equation}
c^{\dag}c = [N]\, , \ \ \ \ c c^{\dag} = [1 + \kappa N] \, ,
\label{cc}
\end{equation}
where the $q$-basic number is defined as
\begin{equation}
[x]=\frac{q^x-q^{-x}}{\qq1}\; . \label{bn}
\end{equation}
In the limit $q \rightarrow 1$, the basic number $[x]$ reduces to
the ordinary number $x$ and all the above relations reduce to the
standard boson and fermion relations.

The transformation from Fock observables to the configuration
space may be accomplished by the replacement (Bargmann holomorphic
representation)\cite{fink}
\begin{equation}
a^\dag\rightarrow x \; , \ \ \ a\rightarrow \dj \; , \label{jd}
\end{equation}
where $\dj$ is the  JD \cite{jack}
\begin{equation}
\dj=\frac{D_x-(D_x)^{-1}}{(q-\q1)\,x} \ ,
\end{equation}
and
\begin{equation}
D_x=q^{x\,\partial_x} \, , \label{dila}
\end{equation}
is the dilatation operator. Its action on an arbitrary real
function $f(x)$ is defined by
\begin{equation}
\dj\,f(x)=\frac{f(q\,x)-f(\q1 x)}{(q-\q1)\,x} \ .
\end{equation}
In contrast to the usual derivative, which measures the rate of
change of the function in terms of an incremental translation of
its argument, the JD measures its rate of change with respect to a
dilatation of its argument by a factor of $q$. The JD plays an
important role in the general $q$-Taylor expansion as discussed in
Ref. \cite{ward} and occurs naturally in $q$-deformed structures;
we will see that it plays a crucial role in the $q$-generalization
of the thermodynamics relations.

The JD satisfies some simple properties which will be useful in
the following. For instance, its action on a monomial $f(x)=x^n$,
where $n\geq0$, is given by
\begin{equation}
\dj\,(a\,x^n)=a\, [n]\,x^{n-1} \ ,\label{7}
\end{equation}
where $a$ is a real constant.

Moreover, it is easy to verify the following basic-version of the
Leibnitz rule
\begin{eqnarray}
\nonumber \dj\Big(f(x)\,g(x)\Big)&=&\dj\,f(x)\,g(\q1 x)+f(q\,x)\,\dj\,g(x) \ ,\\
&=&\dj\,f(x)\,g(q\,x)+f(\q1 x)\,\dj\,g(x) \ .\label{leib}
\end{eqnarray}
In addition the following property holds
\begin{eqnarray}
{\cal D}_{ax}\,f(x)=\frac{1}{a}\,\dj f(x) \ .\label{chain}
\end{eqnarray}

In order to formulate a self-consistent $q$-deformed theory, the standard integral
must be generalized to the basic-integral defined, for $0<q<1$ in
the interval $[0,a]$, as \cite{bona,exton}
\begin{equation}
\int_0^a f(x)\, d_q x=a\,(\q1-q)\, \sum_{n=0}^\infty q^{2n+1}\,
f(q^{2n+1}\,a) \, ,
\end{equation}
while in the interval $[0,\infty)$
\begin{equation}
\int_0^\infty f(x)\, d_q x=(\q1-q)\, \sum_{n=-\infty}^\infty
q^{2n+1}\, f(q^{2n+1}) \, .
\end{equation}
The indefinite $q$-integral is defined as
\begin{equation}
\int f(x)\, d_q x=(\q1-q)\, \sum_{n=0}^\infty q^{2n+1}\,x\,
f(q^{2n+1}\,x)+{\rm constant} \, .
\end{equation}
One can easily see that the $q$-integral approaches the Riemann
integral as $q\rightarrow 1$ and also that $q$-differentiation and
$q$-integration are inverse of each other, thus
\begin{equation}
\dj\int_0^x f(t)\, d_q t=f(x) \, , \ \ \ \int_0^a \dj f(x)\, d_q
x=f(a)-f(0) \, ,
\end{equation}
where the second identity occurs when the function $f(x)$ is
$q$-regular at zero, i.e.
\begin{equation}
\lim_{n\rightarrow \infty} f(xq^n)=f(0) \, .
\end{equation}
By using the deformed Leibnitz rule of Eq.(\ref{leib}), analogous
formulas for integration by parts may easily be deduced as
\begin{eqnarray}
\int_0^a f(qx)\, \dj g(x)\, d_q
x=f(x)\,g(x)\vert^{x=a}_{x=0}-\int_0^a \dj f(x)\, g(\q1 x) \, d_q
x \,  , \\
\int_0^a f(\q1x)\, \dj g(x)\, d_q
x=f(x)\,g(x)\vert^{x=a}_{x=0}-\int_0^a \dj f(x)\, g(q x) \, d_q x
\,  .
\end{eqnarray}

From the above relations, as also pointed out in Ref.
\cite{erzan,sornette}, it appears evident that JD and $q$-calculus
provides a custom made formalism in which to express scaling
relations. When $x$ is taken as the distance from a critical
point, JD thus quantifies the discrete self-similarity of the
function $f(x)$ in the vicinity of the critical point and can be
identified with the generator of fractal and multifractal sets
with discrete dilatation symmetries.

\section{Thermal average and mean occupation number}

Let us start from the following Hamiltonian of non-interacting
$q$-deformed oscillators (fermions or bosons)
\begin{equation}\label{42}
H=\sum_i  (E_i-\mu)\,N_i\, ,
\end{equation}
where $\mu$ is the chemical potential and $E_i$ is the kinetic
energy in the state $i$ with the number operator $N_i$. It is
important to recognize that the latter Hamiltonian, in spite of
the appearance, does include deformation (since the number
operator is deformed by means of Eq.(\ref{cc})), as will become
evident from the form of the average occupation number.

Thermal average of an observable can be computed by following the
usual prescription of quantum mechanics, as follows
\begin{equation}
\langle {\cal O}\rangle=Tr \left (\rho\, {\cal O} \right )\; ,
\end{equation}
where $\rho$ is the density operator and $\cal Z$ is the grand
canonical partition function defined as
\begin{equation}
\rho=\frac{e^{-\beta H}}{\cal Z}\; , \ \ \ \ \ \ {\cal Z}=Tr \left
( e^{-\beta H} \right )\; , \label{pf}
\end{equation}
and $\beta = 1/T$ (hence forward we shall set Boltzmann constant
to unity). We observe that the structure of the density matrix
$\rho = e^{-\beta H}$ and the thermal average are undeformed. As a
consequence, the structure of the partition function is also
unchanged. This is a very common assumption in the literature of
deformed thermodynamics associate to the quantum $q$-deformed
algebra \cite{chai,lee,song,hsu,su}. We emphasize that this is not
a trivial assumption because its validity  implicitly amounts to
an unmodified structure of the Boltzmann-Gibbs entropy, $S=\log\,
W$, where $W$ stands for the number of states of the system
corresponding to the set of occupation numbers $\{ n_i \}$.
Obviously the number $W$ is modified in the $q$-deformed case
\cite{pre2000}. It may be pointed out that in the subject
literature, statistical generalizations are present, such as the
so-called nonextensive thermostatistics or superstatistics with a
completely different origin
\cite{tsallis,book2,varie,beck,abe2,pla02,ditoro}. We like to
stress that in this paper we are dealing with a many body
statistical theory of particles that obey an intermediate behavior
between fermions and bosons, corresponding to a deformed algebra
of the creation and annihilation operators. The deformation
contained into the quantum algebra is reflected into the quantum
statistical behavior. The other way round is realized in other
statistical generalizations present in literature where the
deformation starts from generalized assumptions in statistical
mechanics altering the classical (or quantum) statistical behavior
of complex systems. For example, in the case of the nonextensive
deformed Tsallis statistics, the structure of the entropy is
deformed via the logarithm function and a deformed (classical)
algebra related to a generalized exponential and logarithm
functions emerges in a natural way \cite{borges2,borges3}.

The above assumptions allow us to calculate the average occupation
number $n_i$ defined by the relation $[n_i]= Tr \left (  e^{-
\beta H} c^{\dag}_ic_i\right )/{\cal Z}\, $. Repeated application
of the algebra of $c, c^{\dag}$ along with the use of the cyclic
property of the trace leads to the result
\begin{equation}
n_{i}=\frac{1}{\qq1} \log\left
(\frac{e^{\eta_i}-\overline{\kappa}^{-1}}{
e^{\eta_i}-\overline{\kappa}}\right) \; , \label{nqi}
\end{equation}
where we have set $\eta_i =\beta (E_i- \mu)$ and
$\overline{\kappa}=\kappa\,q^\kappa$ (with the property
$\kappa^{-1}=\kappa$).

 Now we need to further express the occupation
number in a useful form. From Eq.(\ref{nqi}), we arrive at the
result in the form of a power series
\begin{eqnarray}
  n=&&\frac{1}{y}
   +\left(\frac{1}{3\,y^3}+\frac{\kappa}{2\,y^2}\right)\,\epsilon^2+
    \left(\frac{1}{3\,y^3}+\frac{\kappa}{2\,y^2}\right)\,\epsilon^3\nonumber\\
&&+\left(\frac{1}{5\,y^5}+\frac{\kappa}{2\,y^4}+\frac{2}{3\,y^3}+\frac{\kappa}{2\,y^2}\right)
    \,\epsilon^4   + \cdots\label{49} \, ,
\end{eqnarray}
where $y= e^{\eta}- \kappa$ and we have taken $\overline{\kappa} =
\kappa (1- \epsilon), \; \epsilon \ll 1$. At this point, if we use
the approximation by retaining only the leading term,  we then
arrive at the form
\begin{equation}\label{50}
    n\approx \frac{1}{e^{\eta}- \kappa}\, .
\end{equation}
More generally, for $\epsilon$ not very small, we have the series
form in Eq.(\ref{49}) describing the various powers of the
deformation parameter $\epsilon$.

The detailed thermodynamic properties stemming from this form,
such as the equation of state, virial expansion etc.  for the
$q$-bosons and $q$-fermions have been studied in Ref.
\cite{RA-PNS1}. It is also observed that the equality of the
specific heats of boson-type and fermion-type intermediate
statistics particles \cite{May} also prevails as shown in Ref.
\cite{RA-PNS1}, if we utilize the approximate forms for the
occupation numbers. This is a very interesting result and may be
true more generally for the exact forms of the occupation numbers
formulated in the present work.

\section{ The occupation number as an infinite continued fraction}

It is possible to obtain an expression for the mean occupation
number in terms of the infinite Continued Fraction (CF). Let us
begin with the series expression which may be expressed
conveniently in the form
\begin{equation}\label{55}
n=\frac{\alpha_1}{y} +\frac{\alpha_2}{y^2}+ \frac{\alpha_3}{y^3} +
\cdots \, ,
\end{equation}
where, as before, we have set $y=e^{\eta}-\kappa$ (for simplicity,
we have dropped the particle index $i$) and the parameters $\alpha_1, \alpha_2$ etc.
are determined from the previous sections, specifically
Eq.(\ref{49}), such as
\begin{eqnarray}\label{56}
    &&\alpha_1=1\,, \ \ \ \alpha_2=\frac{\kappa}{2}\,
    (\epsilon^2+\epsilon^3+\epsilon^4+\cdots) \, ,
    \nonumber\\
&&\alpha_3=\frac{1}{3}\,(\epsilon^2+\epsilon^3+2\,\epsilon^4+\cdots)
\, ,
\end{eqnarray}
etc. by combining terms containing various powers of $\epsilon$ in
Eq.(\ref{49}).  There is a standard method by which this infinite
series can be put in the form of CF. The method of determining the
CF form of a function given by an infinite series is well-known in
the literature \cite{Wall,Andrews}.

We shall briefly summarize the procedure here. The general
continued fraction of order $r$ is of the form
\begin{equation}\label{60}
    C_r=  b_0+ {a_1\over \displaystyle b_1 + {\strut a_2 \over
    \displaystyle
     b_2 + {\strut a_3 \over \displaystyle b_3 + \cdots }}}\, ,
\end{equation}
where the constants $b_0, b_1, \cdots \; , a_0, a_1, \cdots $ can
be determined by a straightforward procedure. \\
The various
convergents are $C_0, C_1, \cdots $ corresponding to
$r=0,1,2,\cdots, \infty$. Accordingly we have
\begin{eqnarray}\label{60a}
    &&C_0=b_0=: A_0/B_0 \, ; \nonumber \\
    &&C_1=b_0 + a_1/b_1= \frac{b_0b_1 + a_1}{b_1}=: \frac{A_1}{B_1}
    \,; \nonumber\\
&&C_2=b_0 + a_1/(b_1 + a_2/b_2)= b_0 + a_1b_2/(b_1 b_2 +
    a_2)=:\frac{A_2}{B_2}\, ,
\end{eqnarray}
etc. The parameters $A_n, B_n$ satisfy the two-term recurrence
relations \cite{Andrews}:
\begin{eqnarray}\label{60b}
    &&A_n=b_n A_{n-1}+ a_n A_{n-2}; \; A_{-1}=1 \, ;\nonumber\\
    &&B_n= b_n B_{n-1} + a_n B_{n-2}; \; B_{-1}=0\, .
\end{eqnarray}
By solving the recurrence relations,  the general CF can be
determined. We may quote two examples of this procedure. The
standard sine series may be expressed in the form of a CF as:
\begin{equation}\label{60c}
    \sin x=   {x\over \displaystyle 1 + {\strut x^2 \over
    \displaystyle
     2\cdot 3 - {\strut x^2 + 2\cdot 3 x^2\over \displaystyle 4 \cdot 5 -x^2 + \cdots }}}\,
     .
\end{equation}
Furthermore, we can also deal with the inverse problem, i.e.,
given the standard series form of the cosine function,
\begin{equation}\label{60d}
    \cos x = 1- x^2/2! + x^4/4! + \cdots \, ,
\end{equation}
we can employ the above procedure and obtain the CF form for the
cosine function as:
\begin{equation}\label{60e}
    \cos x=   {1\over \displaystyle 1 + {\strut x^2 \over
    \displaystyle
     2\cdot 1 - {\strut x^2 + 2 x^2\over \displaystyle 4 \cdot 3 -x^2 + \cdots }}}\,
     .
\end{equation}
 Employing this procedure for our present problem, after some
algebra, the final result can be expressed by accordingly
obtaining  the various convergents (approximants):
\begin{eqnarray}\label{57}
n_1&=& \frac{\alpha_1}{y}\, , \\
\label{58} n_2&=& -\frac{\alpha_2 \, y}{\alpha_1 \, y + \alpha_2}
\,
, \\
\label{59} n_3&=& -\frac{\alpha_3\, y}{\alpha_2 \, y + \alpha_3}
\, ,
\end{eqnarray}
etc.\\
%
%
In the literature on CF, the convergents, which may be obtained in
a straightforward manner after some algebra, play an important
role.

The general form of the CF is given by the form $C_r$ as in
Eq.(\ref{60}) and the procedure  can be extended to many
convergents. The meaning of the convergent or the approximant is
evident.

Now the question which might arise is: what is the advantage of
CF? Other than the elegant mathematical form, we remark that there
is a distinct advantage. The Pade approximant is a well-known
application. Moreover there is a theorem \cite{Wall}, involving
the convergents $n_1, n_2, n_3 \cdots$ which may be stated as:
\begin{equation}\label{61a}
    n_1 < n_3 < \cdots <n   \;\;\;{\rm and}\;\;\;
    n_2 > n_4 . \cdots > n\, .
\end{equation}
 This immediately provides a
clarifying definition of successive approximations i.e., the above
inequality tells us how to obtain successive approximations of the
quantity $n$. Indeed, the above tells us immediately that the
exact form of $n$ lies between $n_1$ and $n_2$, hence its
importance. We can thus establish that the exact $n$ is bigger
than the first convergent $n_1= \alpha_1 / y $ but smaller than
$n_2$ obtained above.

\section{Thermodynamics of $q$-deformed bosons and fermions}

In Ref.~\cite{pre2000}, we have shown that the entire structure of
thermodynamics is preserved if the ordinary derivatives are replaced
by the use of an appropriate JD
\begin{equation}
\frac{\partial}{\partial z} \Longrightarrow {\cal D}^{(q)}_z \; .
\end{equation}
Consequently, posing the fugacity $z=e^{\beta\mu}$, the number of
particles in the $q$-deformed theory can be derived from the
relation
\begin{equation}
N=z \; {\cal D}^{(q)}_z \ln {\cal Z}\equiv \sum_i n_i \; ,
\label{numq}
\end{equation}
where $n_i$ is the mean occupation number expressed in
Eq.(\ref{nqi}).

The usual Leibniz chain rule is ruled out for the JD and therefore
derivatives encountered in thermodynamics must be modified as
follows. First we observe that the JD applies only with respect to
the variable in the exponential form such as $z=e^{\beta \mu}$ or
$y_i=e^{-\beta \epsilon_i}$. Therefore for the $q$-deformed case,
any thermodynamic derivative of functions which depend on $z$ or
$y_i$ must be transformed to derivatives in one of these variables
by using the ordinary chain rule and then evaluating the JD with
respect to the exponential variable.  For instance, in the  case
of the internal energy in the $q$-deformed case, we can write this
prescription explicitly as
\begin{equation}
U=-\left. \frac{\partial}{\partial\beta} \ln {\cal Z} \right |_z=
\kappa\sum_i \frac{\partial y_i}{\partial\beta} \, {\cal
D}^{(q)}_{y_i}\ln(1-\kappa z\,y_i) \; . \label{intq}
\end{equation}
In this case we obtain the correct form of the internal energy
\begin{equation}
U=\sum_i \epsilon_i \, \nq\; , \label{un}
\end{equation}
where $n_i$ is the mean occupation number expressed in
Eq.(\ref{nqi}).

In the thermodynamic limit, for a large volume $V$ and a large
number of particles, the sum over states can be replaced by the
integral. However, as previous discussed, in a $q$-deformed theory
the standard integral should be consistently generalized to the
$q$-integral, inverse operator of the JD. In this manner, we
extent our previous formulation by employing the $q$-integral
operator and, following the above prescriptions, we have
\begin{equation}
\sum_i f(u_i) \, \Longrightarrow
\,I_q=g_\kappa\,\frac{V}{(2\pi)^3} \, \int\!\!\! \, f[u(k)] \,
d_qk_x \,\, d_qk_y \,\, d_qk_z\, , \label{intk}
\end{equation}
where $g_\kappa$ is the spin degeneracy factor, $u(k)=\beta\,
\hbar^2 k^2/2m$ and satisfies the constraint:
$k^2=k_x^2+k_y^2+k_z^2$ \cite{wachter}. By taking into account
the rules related to changing the variable of $q$-integration
\cite{exton}, we have verified that for $0.6<q<1.4$ the above
integration can be well approximately expressed as
\begin{equation}
I_q \approx g_\kappa\,\frac{2}{\sqrt{\pi}}\,\frac{V}{\lambda^3} \,
\frac{2}{q+\q1}\, \int_0^\infty \!\!\! \, f(u)\, d_Qu \, ,
\label{int2dq}
\end{equation}
where $Q=q^2$ (a change of variable $u=\beta\, \hbar^2 k^2/2m$
also involves a corresponding change of base) and $\lambda =
h/(2\pi m T)^{1/2}$ is the thermal wavelength. Therefore, in the
thermodynamic limit, Eq.(\ref{numq}) and Eq.(\ref{intq}),
respectively, becomes \bee
&&N_{\kappa}(T,z)=g_\kappa\,\frac{V}{\lambda^3}
\, h_{3/2}^\kappa(z,q) \, ,\label{nq}\\
&&U_{\kappa}(T,z)=g_\kappa\,\frac{3}{2}\,\frac{V}{\beta\,\lambda^3}
\, h_{5/2}^\kappa(z,q)\, , \label{uq}\eee where we have defined
the $q$-deformed $h_n^\kappa(z,q)$ as
\begin{equation}
h_n^\kappa(z,q)=\frac{1}{\Gamma (n)} \int_0^\infty \!\!\!
 \frac{u^{n-1}}{\qq1} \ln\left (\,\frac{z^{-1}e^u-\kappa \,
q^{-\kappa}} { z^{-1}e^u-\kappa \, q^\kappa} \right) \, d_Qu \; .
\label{hn}
\end{equation}

It must be stressed that the above equation is quite different
from the definition of the generalized function introduced in
Eq.(21) in our earlier work \cite{pre2002}. This is an important
notion in our present work. It should also be noted that, to the
best of our knowledge, this is the first time that $q$-integrals
are numerically employed in thermostatistics calculations. In the
limit $q\rightarrow 1$, the deformed $h_n^\kappa(z,q)$ functions
reduce to the standard $h_n^\kappa(z)$ for bosons and fermions.

As in the undeformed boson case, we need to set the range of the
$q$-boson fugacity $z_B$ which will correspond to non-negative
occupation number. In the case of $q$-bosons we see that the
condition is $z_B< 1/q$ for $q>1$ and $z_B<1$ for $q<1$. Moreover,
it should be pointed out that we also have to require the
existence of the JD of the mean occupation number which is
encountered in the calculation of thermodynamic quantities such as
the specific heat and this changes the upper bound of the fugacity
$z_B$. In the following, we thus will require the condition $z_B<
z_q$, where we have defined

\begin{equation}
z_q=\cases{ q^{-2} &if $q>1$ ; \cr  q^2 &if $q<1$ .\cr} \label{zq}
\end{equation}

\section{Specific heat of boson and fermion systems}

We are now able to calculate the specific heat of the $q$-boson
and $q$-fermion gas, starting from the thermodynamic definition
\begin{equation}
C_v=\left. \frac{\partial U}{\partial T}\right|_{V,N}\;
.\label{cvt}
\end{equation}

Carrying out the JD prescription, described earlier,
Eq.(\ref{cvt}) in the $q$-deformed theory can be written as
\begin{equation}
C_v=-\beta^2 \sum_i \, \epsilon_i\,\frac{\partial
\gamma_i}{\partial \beta}\; \frac{1}{q-\q1} {\cal
D}_{\gamma_i}^{(q)}\ln \left ( \frac{1-\kappa\,
q^{-\kappa}\,\gamma_i}{1-\kappa\,q^\kappa\,\gamma_i}  \right ) \;
, \label{cvq}
\end{equation}
where $\gamma_i= z \, e^{- \beta \epsilon_i}$ and
\begin{equation}
\frac{\partial \gamma_i}{\partial \beta}=\left(
\frac{1}{z}\frac{\partial
z}{\partial\beta}-\epsilon_i\right)\,\gamma_i\, .
\end{equation}

For this purpose we first need, therefore, the derivative of the
fugacity with respect to $T$ (or $\beta$), keeping $V$ and $N$
constant.  Accordingly, we observe that the following identity
holds (since the number of particles is kept constant)
\begin{equation}
\frac{\partial}{\partial \beta}\sum_i \ln \left ( \frac{1-\kappa\,
q^{-\kappa}\,\gamma_i}{1-\kappa\,q^\kappa\,\gamma_i}  \right ) = 0
\; .
\end{equation}
In accordance with  the JD recipe about the thermodynamical
relations, the above equation can be written as
\begin{equation}
\sum_i \, \frac{\partial \gamma_i}{\partial \beta}\, {\cal
D}_{\gamma_i}^{(q)}\ln \left ( \frac{1-\kappa\,
q^{-\kappa}\,\gamma_i}{1-\kappa\,q^\kappa\,\gamma_i}  \right )) =
0\; .
\end{equation}
Evaluating in the thermodynamical limit ($V\rightarrow\infty$) and
by using the definition in Eq.(\ref{hn}), we obtain
\begin{equation}
\left. \frac{1}{z}\, \frac{\partial
z}{\partial\beta}\right|_{V,N}= \frac{3}{2}\,\frac{1}{\beta} \;
\frac{{\cal D}^{(q)}_z h_{5/2}^\kappa (z,q)}{ {\cal D}^{(q)}_z
h_{3/2}^\kappa (z,q)} \; . \label{dzb}
\end{equation}

By using the above relation in Eq.(\ref{cvq}), we obtain the
specific heat for a system of bosons and fermions at fixed $T$ and
$N$
\begin{equation}
\frac{C_v\,\lambda^3}{V}=g_\kappa \,\left\{ \frac{15}{4} z\, {\cal
D}^{(q)}_z h_{7/2}^\kappa (z,q) -\frac{9}{4}\,z\, \frac{({\cal
D}^{(q)}_z h_{5/2}^\kappa (z,q))^2}{ {\cal D}^{(q)}_z
h_{3/2}^\kappa (z,q)}\right\} \; . \label{cvfinal}
\end{equation}

In Fig. 1 and 2 we display the specific heat
$C_v\,\lambda^3/(g_\kappa\,V)$ as a function of the fugacity for
bosons (above the critical point of boson condensation) and
fermions, respectively (note that the range of meaningful
fugacities $z_B$, for boson gas, is limited by the condition
(\ref{zq})) for different values of $q$. We remember that we are
employing the symmetric $q \leftrightarrow q^{-1}$ deformed
quantum algebra, therefore the displayed graphs are identical for
the transformation $q \leftrightarrow q^{-1}$ (more explicitly:
the long dashed curves in Fig. 1 and Fig. 2 stand for $q=0.8$ and
$q=1/0.8$, the short dashed curves stand for $q=0.7$ and
$q=1/0.7$). Finally, let us observe that the modification of the
specific heat increasing with the value of the deformation
parameter $q$ becomes very remarkable in the fermion case.

\begin{figure}[bt]
\resizebox{0.8\textwidth}{!}{%
\includegraphics{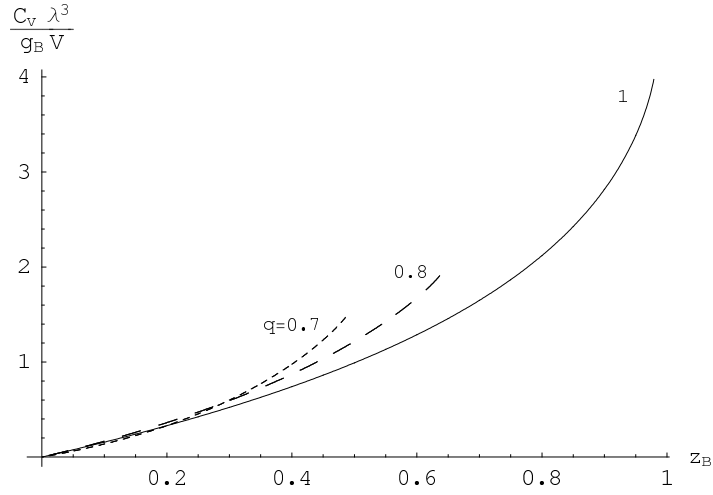}}
\caption{The specific heat $C_v\,\lambda^3/(g_B V)$ for bosons as
a function of fugacity $z_B$ for different values of $q$. Same
values are obtained for the transformation $q \leftrightarrow
q^{-1}$.}
\end{figure}

\begin{figure}[bt]
\resizebox{0.8\textwidth}{!}{%
\includegraphics{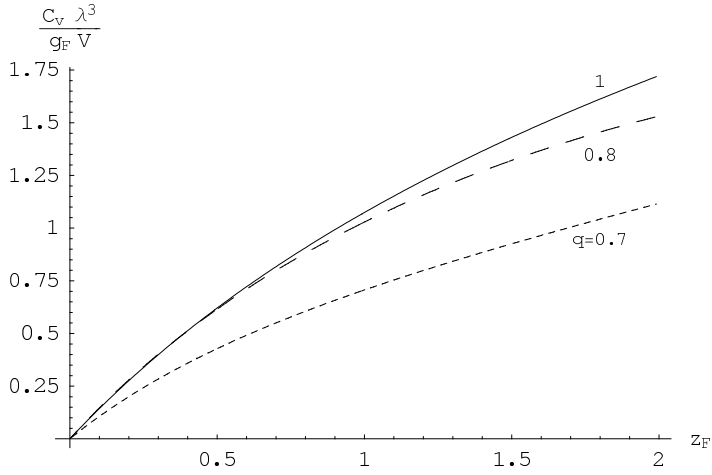}}
\caption{The specific heat $C_v\,\lambda^3/(g_F V)$ for a fermion
gas as a function of fugacity $z_F$ for different values of $q$.
Same values are obtained for the transformation $q \leftrightarrow
q^{-1}$.}
\end{figure}

\section{Conclusion}

In this paper, we have investigated the structure of symmetric $q
\leftrightarrow q^{-1}$ deformed quantum thermostatistics by
working, consistently in the framework of the $q$-calcu\-lus, with
the use of the JD and the $q$-integration. We have shown that the
entire structure of thermodynamics is preserved if the ordinary
derivatives and integrals are replaced by the JD and $q$-integral,
respectively. This prescription indeed gives us a recipe to obtain
the fundamental thermodynamic functions such the mean occupation
number, the specific heat etc. We may point out that this is a
relevant premise of our original approach since, to the best of
our knowledge, this is the first time $q$-derivatives and
$q$-integrals are consistently employed in thermodynamical
investigations.

Our formulation of the mean occupation number and other
thermodynamic parameters in terms of the infinite continued
fraction is a new feature, not known in the literature. Its
importance stems from the possibility of approximations i.e., the
validity of approximations in the theory. The behavior different
from the undeformed quantum theory can be dealt with in the
statistical behavior of a complex system, intrinsically contained
in the $q$-deformation, whose underlying dynamics is spanned in
many body interactions and other long time memory effects. This
aspect has been outlined in many papers in the recent literature.

The different behavior from the undeformed quantum theory can be
dealt with in the statistical behavior of a complex systems,
intrinsically contained in $q$-deformation, whose underlying
dynamics is spanned in many-body interactions and/or long-time
memory effects. This aspect is  outlined in several papers. For
example in Ref.\cite{svira} it has been shown that $q$-deformation
plays a significant role in understanding higher-order effects in
many-body nuclear interactions. Moreover, the strong effects on
the deformation, that we have found especially in the $q$-fermion
specific heat, could be connected to an intrinsic presence of
complex many-body effective interactions on $q$-deformation
theory. In this context, it appears relevant to observe that
nonanalytic temperature behavior of the specific heat of Fermi
liquid can be explained within two dimensional interactions beyond
the weak-coupling limit \cite{chubo}.


\begin{thebibliography}{}

\bibitem{spin}
W. Pauli, Ann. Inst. Henri Poincar\'e {\bf 6}, 137 (1936); W.
Pauli, Phys. Rev. {\bf 58}, 716 (1940); S. Weinberg, Phys. Rev.
{\bf 133}, B1318 (1964).
\bibitem{berry}
M.V. Berry, J.M. Robbins, Proc. R. Soc. Lond. A {\bf 453} 1771
(1997); J.M. Leinass, J. Myrheim, Nuovo Cimento B {\bf 37}, 1
(1977).

\bibitem{gentile}
G. Gentile, Nuovo Cimento {\bf 17}, 493 (1940).
\bibitem{green}
H.S. Green, Phys. Rev. {\bf 90}, 270 (1953).
\bibitem{greenb1}
O.W. Greenberg, R.N. Mohapatra, Phys. Rev. Lett. {\bf 59}, 2507
(1987); Phys. Rev. Lett. {\bf 61}, 1432 (1988); Phys. Rev. Lett.
{\bf 62}, 712 (1989)
\bibitem{greenb2}
O.W. Greenberg, Phys. Rev. Lett. {\bf 64}, 705 (1990); Phys. Rev.
D {\bf 43}, 4111 (1991).
\bibitem{vieira}
M.C. de Sousa Vieira, C. Tsallis, J. Stat. Phys. {\bf 48}, 97
(1987).
\bibitem{poli}
A.P. Polychronakos, Nucl. Phys. B {\bf 474}, 529 (1996).
\bibitem{levine}
R.Y. Levine, Y. Tomazawa, Phys. Lett. B {\bf 128}, 189 (1983).


\bibitem{reines}
F. Reines, H.W. Sobel, Phys. Rev. Lett. {\bf 32}, 954 (1974).
\bibitem{hilborn}
R.C. Hilborn, C.L. Yuca, Phys. Rev. Lett. {\bf 76}, 2844 (1996).
\bibitem{modugno}
G. Modugno, M. Inguscio, G.M. Tino, Phys. Rev. Lett. {\bf 81},
4790 (1998).
\bibitem{javorsek}
D. Javorsek {\it et al.}, Phys. Rev. Lett. {\bf 87}, 231804
(2001).
\bibitem{vip}
S. Bartalucci {\it et al.}, Phys. Lett. B {\bf 641}, 18 (2006); E.
Milotti {\it et al.}, Int. J. Mod. Phys. A {\bf 22}, 242 (2007);
D. Pietreanu {\it et al.}, Int. J. Mod. Phys. A {\bf 24}, 190
(2009).

\bibitem{hilborn2}
C.C. Gerry, R.C. Hilborn, Phys. Rev. A {\bf 55}, 4126 (1997).
\bibitem{green_hilb} O.W. Greenberg, R.C.
Hilborn, Phys. Rev. Lett. {\bf 83}, 4460 (1999).
\bibitem{baron}
E. Baron, R.N. Mohapatra, V.L. Teplitz, Phys. Rev. D {\bf 59},
036003 (1999).
\bibitem{strominger}
A. Strominger, Phys. Rev. Lett. {\bf 71}, 3397 (1993).
\bibitem{youm}
D. Youm, Phys. Rev. D {\bf 62}, 095009 (2000).



\bibitem{sky}
E.K. Skylyanin, Func. Analysis Appl. {\bf 16}, 262 (1982).
\bibitem{kulish}
P.P. Kulish, N.Y. Reshetikhin, Jour. Sov. Math., {\bf 23}, 2435
(1983).


\bibitem{bie}
L. Biedenharn, J. Phys. A {\bf 22}, L873 (1989).
\bibitem{mac}
A. Macfarlane, J. Phys. A {\bf 22}, 4581 (1989).

\bibitem{heine}
E. Heine, J. reine angew. Math. {\bf 32}, 210 (1846); {\bf 34},
285 (1847);  {\it Handbuch der Kugelfunctionen, Theorie und
Anwendungen}, Vol. 1, (Reimer, Berlin, 1878).
\bibitem{jack}
F. Jackson, Mess. Math. {\bf 38}, 57 (1909).
\bibitem{gasper}
G. Gasper, M. Rahman, {\it Basic Hypergeometric Series},
Encyclopedia of mathematics and its applications (Cambridge
Univeristy Press, 1990).
\bibitem{cele1}
E. Celeghini {\it et al.}, Ann. Phys. {\bf 241}, 50 (1995).
\bibitem{fink}
R. J. Finkelstein, Int. J. Mod. Phys. A {\bf 13}, 1795 (1998).
\bibitem{abe}
S. Abe, Phys. Lett. A {\bf 224}, 326 (1997); G. Kaniadakis, A.
Lavagno, P. Quarati, Phys. Lett. A {\bf 227}, 227 (1997); Nucl.
Phys. B {\bf 466}, 527 (1996).
\bibitem{borges1}
E.P. Borges, J. Phys. A: Math. Theor. {\bf 31}, 5281 (1998).

\bibitem{erzan}
A. Erzan, J.-P. Eckmann, Phys. Rev. Lett. {\bf 78}, 3245 (1997);
A. Erzan, Phys. Lett. A {\bf 225}, 235 (1997).
\bibitem{lava08}
A. Lavagno, J. Phys. A: Math. Theor. {\bf 41} 244014 (2008); A.
Lavagno, P. Narayana Swamy, Int. J. Mod. Phys. B {\bf 23}, 235
(2009).

\bibitem{pre2000}
A. Lavagno and P. Narayana Swamy, Phys. Rev. E {\bf 61}, 1218
(2000).
\bibitem{pre2002}
A. Lavagno and P. Narayana Swamy, Phys. Rev. E {\bf 65}, 036101
(2002).
\bibitem{jpa2007}
A. Lavagno, A.M. Scarfone, P. Narayana Swamy, Eur. Phys. J. C {\bf
47}, 253 (2006); A. Lavagno, A.M. Scarfone, P. Narayana Swamy, J.
Phys. A: Math. Theor. {\bf 40}, 8635 (2007).


\bibitem{ng}
Y.J. Ng, J. Phys. A {\bf 23}, 1023 (1990).
\bibitem{chai}
M. Chaichian, R. Gonzalez Felipe, C. Montonen, J. Phys. A {\bf
26}, 4017 (1993).
\bibitem{lee}
C.R. Lee, J.P. Yu, Phys. Lett. A {\bf 150}, 63 (1990).
\bibitem{song}
H.S. Song, S.X. Ding and I. An, J. Phys. A {\bf 26}, 5197 (1993).

\bibitem{ward}
M. Ward, Am. J. Math., {\bf 33}, 255 (1936).

\bibitem{bona}
D. Bonatsos and C. Daskaloyannis, Progr. Part. Nucl. Phys. {\bf
43}, 537 (1999).
\bibitem{exton}
H. Exton, {\it $q$-Hypergeometric Functions and Applications}
(John Wiley and Sons, New York 1983).

\bibitem{sornette}
W-X Zhou and D. Sornette: {\em Phys. Rev. E\/} {\bf 66}, 046111
(2002).

\bibitem{hsu}
R.R. Hsu, C.R. Lee, Phys. Lett. A {\bf 180}, 314 (1993).
\bibitem{su}
G. Su, M. Ge, Phys. Lett. A {\bf 173}, 17 (1993).

\bibitem{tsallis}
C. Tsallis, J. Stat. Phys. {\bf 52}, 479 (1988); M. Gell-Mann and
C. Tsallis, eds., {\it Nonextensive Entropy: Interdisciplinary
Applications} (Oxford University Press, New York, 2004).
\bibitem{book2}
C. Tsallis, {\it Introduction to Nonextensive Statistical
Mechanics} (Springer-Verlag, New York, 2009).
\bibitem{varie}
U. Tirnakli, C. Beck, C. Tsallis, Phys. Rev. E {\bf 75}, 040106
(2007); C. Tsallis {\it et al.}, Physica A {\bf 381}, 143 (2007);
A. Pluchino, A. Rapisarda, C. Tsallis, Physica A {\bf 387}, 3121
(2008); F. Caruso, C. Tsallis, Phys. Rev. E {\bf 78}, 021102
(2008); U. Tirnakli, C. Tsallis, C. Beck, Phys. Rev. E {\bf 79},
056209 (2009); U. Tirnakli, D.F. Torres, Physica A 268 (1999) 225.
\bibitem{beck}
C. Beck, E.G.D. Cohen, Physica A {\bf 322}, 267 (2003); C. Beck,
Physica A {\bf 342}, 459 (2004); C. Beck, Phys. Rev. Lett. {\bf
98}, 064502 (2007); E. Van der Straeten, C. Beck, Phys. Rev. E
{\bf 78}, 051101 (2008).

\bibitem{abe2}
A. Lavagno, P. Quarati, Phys. Lett. B {\bf 498}, 47 (2001); Nucl.
Phys. B [PS] {\bf 87}, 209 (2000).
\bibitem{pla02}
A. Lavagno, Phys. Lett. A {\bf 301}, 13 (2002); A. Lavagno,
Physica A {\bf 305}, 238 (2002);
\bibitem{ditoro}
M. Di Toro {\it et al.}, Nucl. Phys. A {\bf 775}, 102 (2006); A.
Drago, A. Lavagno, P. Parenti, Ap. J. {\bf 659}, 1519 (2007); W.M.
Alberico, A. Lavagno, Eur. Phys. J. A {\bf 40}, 313 (2009).

\bibitem{borges2}
E.P. Borges, Physica A {\bf 340}, 95 (2004).
\bibitem{borges3}
P.G.S. Cardoso {\it et al.}, J. Math. Phys. {\bf 49}, 093509
(2008).


\bibitem{RA-PNS1}
R. Acharya, P. Narayana Swamy, J. Phys. A: Math. Gen. {\bf 27},
7247 (1994) and references cited therein

\bibitem{May}
R.M. May , {\it Phys. Rev.} {\bf 135}, A1515 (1964).

\bibitem{Wall}
H.S. Wall, {\it Analytic theory of continued fractions}
(Princeton: Van Nostrand Company) (1948).

\bibitem{Andrews}
G. Andrews {\it et al}, {\it Special Functions}, Cambridge:
Cambridge University Press  (1999).


\bibitem{wachter}
H. Wachter, Eur. Phys. J. C {\bf 32}, 281 (2004).

\bibitem{svira}
K.D. Sviratcheva, C. Bahri, A.I. Georgieva, J.P. Draayer, Phys.
Rev. Lett. {\bf 93}, 152501 (2004).

\bibitem{chubo}
A.V. Chubukov, D.L. Maslov, S. Gangadharaiah, L.I. Glazman, Phys.
Rev. Lett. {\bf 95}, 026402 (2005).







\end{thebibliography}
\end{document}